\documentclass{mn2e}

 \usepackage[dvips]{epsfig}
 \usepackage{amsfonts,amsbsy}

\title [{\it I\/} band flaring in SAX J1808.4-3658]
{A transient {\it I\/} band excess in the optical spectrum of the accreting millisecond pulsar SAX J1808.4--3658}
\author [J.G.Greenhill {\it et al.}]       
{ J. G. Greenhill$^{1}$, A. B. Giles$^{1,2}$ and C Coutures$^{3}$\\
      $^{1}$ School of Mathematics and Physics, University of Tasmania, 
             Private bag 37, Hobart, Tasmania 7001, Australia \\
      $^{2}$ Spurion Technology Pty. Ltd., 200 Mt. Rumney Road, Mt. Rumney, 
             Tasmania 7170, Australia \\
      $^{3}$ CEA/Saclay, 91191 Gif-sur-Yvette cedex, France } 

\date{Accepted 10 May 2006}

\pagerange{\pageref{firstpage}--\pageref{lastpage}}
\pubyear{2005}

\begin{document}
\maketitle

\label{firstpage}

\begin{abstract}
The optical counterpart of the transient, millisecond X-ray pulsar SAX J1808.4-3658
was observed in four colours ({\it BVRI\/}) for five weeks during the 2005 June-July outburst.
The optical fluxes declined by $\sim 2$ magnitudes during the first 16 days and then commenced 
quasi-periodic secondary outbursts, with time-scales of several days, similar to those seen in 
2000 and 2002. The broadband spectra derived from these measurements were 
generally consistent with emission from an X-ray heated accretion disc. During the first 16 days 
decline in intensity the spectrum became redder. We suggest that the primary
outburst was initiated by a viscosity change driven instability in the inner disc and 
note the contrast with another accreting 
millisecond pulsar, XTE J0929--314, for which the spectrum becomes bluer during the decline. 
On the night of 2005 June 5 (HJD 2453527) the {\it I\/} band flux was $\sim 0.45$ magnitudes brighter 
than on the preceding or following nights whereas the {\it BV\/} \& {\it R\/} bands showed no obvious 
enhancement. A Type I X-ray burst was detected by the {\it RXTE} spacecraft during this {\it I\/} band 
integration. It seems unlikely that reprocessed radiation from the burst was sufficient to explain 
the observed increase. We suggest that a major part of the {\it I\/} band excess was due to
synchrotron emission triggered by the X-ray burst.  Several other significant short duration changes 
in {\it V-I\/} were detected. One occurred at about HJD 2453546 in the early phase of the first 
secondary outburst and may be due to a mass transfer instability or to another synchrotron emission event.

\end{abstract}

\section{Introduction} 
SAX J1808.4--3658 was the first transient millisecond X-ray pulsar to be 
discovered and has been studied  extensively at 
all wavelengths (Chakrabarty \& Morgan 1998; in 't Zand et al. 1998; Wijnands \& van der Klis 1998; Giles, Hill \& Greenhill 1999;  Wachter et al. 2000; 
Wang et al. 2001; Wijnands et al. 2001;  Homer et al. 2002; Markwardt, Miller 
\& Wijnands 2002;  Chakrabarty et al. 2003; Wijnands et al. 2003). 
Study of these systems is providing important 
information on the evolutionary path by which a conventional LMXB system 
becomes a millisecond radio pulsar (Campana et al. 2004; Bogdanov, Grindlay \& van den Berg 2005; Wijnands 2005). 

The optical spectrum of SAX J1808.4--3658 during the 1998 outburst was found by Wang et al. (2001) to be 
consistent with emission from an X-ray irradiated accretion disc although there
was a clear IR ({{\it JHK\/})excess on one occasion. It is possible that this 
extended into the optical bands (Giles et al. 2005). A 2 hour orbital period modulation in {\it V\/} 
with amplitude $ 0.12 \pm 0.02$ magnitudes peak to peak  was detected
during the 1998 outburst (Giles et al. 1999).  The phase was consistent with X-ray
heating of the companion. A much deeper orbital modulation, too large to be generated by X-ray
heating, was present in the quiescent state (Campana et al. 2004). They suggested that heating of the 
companion by a pulsar generated relativistic particle wind was responsible.

SAX J1808.4--3658 is the only member of the 7 known accreting millisecond pulsars known to 
undergo extended low level activity states. These follow the initial outburst and are characterised 
by erratic, large amplitude variability on time-scales of hours to days. Repeated secondary X-ray outburts 
with durations of several days are sometimes accompanied by optical outbursts (Wijnands, 2004). Disc 
instability and mass transfer instability models have been suggested as possible mechanisms 
(Wijnands et al. 2001). 

Recently it has become apparent that synchrotron emission, probably from matter 
flowing out of the system via bipolar jets, makes a highly variable 
contribution to radio and IR emission from many different classes of X-ray 
binaries (Fender 2003). In some cases this emission may extend into the 
optical region (Hynes et al. 2000). Krauss et al. (2005) detected
an {\it I\/} band excess during the discovery outburst of the accretion-powered 
millisecond pulsar XTE J1814--318 and Giles et al. (2005) found evidence
of a variable {\it I\/} band excess from another accretion-powered 
millisecond pulsar XTE J0929--314. 

In this paper we describe the broad-band optical spectra of SAX J1808.4--3658 
measured in 4 colours during the 2005 June outburst.

\section{Observations}
All the observations described in this paper were made using the 1-m 
telescope at the University of Tasmania, Mt. Canopus Observatory. The CCD 
camera, its operating software (CICADA), the image reduction and analysis 
tools (MIDAS and DoPHOT) were identical to that described in Giles et al. 
(1999). The data were reduced using the PLANET microlensing collaboration 
pipeline QUYLLUR WASI. 
The CCD camera contains an SITe 512 x 512 pixel
thinned back illuminated chip with an image scale of 
$0.434 \arcsec $ pixel$^{-1}$. Cousins standard {\it BVRI\/} filters (Bessell 1990) 
were used for the observations. The magnitudes of three stable local reference stars 
(Fig. 1) were determined using a 
sequence of observations of three standard stars within the E7 field of  
Graham (1982). Both source and standard star fields were observed at 
virtually identical airmasses (1.01--1.03) and experience with this camera has shown that 
differential colour corrections are generally negligible.
From these observations we derived 
the magnitudes of three local secondary standards close to SAX J1808.4--3658 
and within the CCD frame. These local standards are marked as stars 1--3 on 
the finder chart in Fig. 1 and we tabulate their derived magnitudes in 
Table 1. The magnitudes for SAX J1808.4--3658 were then obtained using 
differential photometry relative to these local secondary standards. 
The complete data set from 2005 June 2 to 
July 6 (HJD 245[3524] -- 245[3558]) is detailed in Table 2 and 
forms the subject of this paper.

\begin{table}
 \caption{The magnitudes of local standard stars 1--3 in Fig. 1.}
 \label{symbols}
 \begin{tabular}{@{}ccccc}
 \hline
 Star No. &  {\it B\/}  &  {\it V\/}  &  {\it R\/}  &  {\it I\/}  \\
 \hline
    1     &   15.60    &   14.76    &   14.23    &   13.74    \\
    2     &   16.89    &   15.31    &   14.50    &   13.71    \\
    3     &   15.93    &   15.18    &   14.71    &   14.34    \\    
  Mean    &  16.02(4)  &  15.06(3)  &   14.46(3) &   13.89(3) \\   
 \hline
 \end{tabular}
 \medskip
\end{table} 

\begin{table*}
\centering
\begin{minipage}{175mm}
\begin{center}
 \bf Table 2. \rm A journal of the observations. 
 \vspace*{0.125cm}
 \label{symbols}
 \begin{tabular}{@{}llrllrllrllr}
 \hline
            &                & Int.    &           &                & Int.    &           &                & Int.    &           &                & Int.     \\ 
  $HJD^{a}$ & {\it I\/} mag. &  (s)    & $HJD^{a}$ & {\it R\/} mag. &  (s)    & $HJD^{a}$ & {\it V\/} mag. &  (s)    & $HJD^{a}$ & {\it B\/} mag. &  (s)     \\ 
 \hline
   24.0918  &  17.12(1)      &  300    &           &                &         &  24.0978  &  17.43(1)      &  300    &           &                &          \\
   25.1991  &  17.15(2)      &  300    &  25.2076  &  17.14(1)      &  300    &  25.2119  &  17.31(1)      &  300    &  25.2183  &  17.29(1)      &  600     \\
   27.1449  &  16.72(1)      &  300    &  27.2305  &  17.19(1)      &  300    &  27.1673  &  17.29(2)      &  300    &  27.2367  &  17.36(1)      &  600     \\
   29.2256  &  17.20(4)      &  300    &  29.2324  &  17.33(2)      &  300    &  29.2371  &  17.45(2)      &  300    &  29.2485  &  17.48(2)      &  600     \\
   31.1353  &  17.25(4)      &  300    &  31.1433  &  17.33(2)      &  300    &  31.1393  &  17.48(2)      &  300    &  31.1492  &  17.60(1)      &  600     \\
   34.1917  &  17.36(3)      &  300    &  34.1960  &  17.56(2)      &  300    &  34.1999  &  17.82(3)      &  300    &  34.2058  &  17.79(2)      &  600     \\
   36.2103  &  17.65(3)      &  300    &  36.2189  &  17.86(2)      &  300    &  36.2146  &  18.03(2)      &  300    &  36.2249  &  18.17(1)      &  600     \\
   36.2811  &  17.63(2)      &  300    &  36.2887  &  17.80(2)      &  300    &  36.2849  &  18.01(2)      &  300    &  36.2944  &  18.04(4)      &  600     \\
   37.1149  &  17.76(2)      &  300    &  37.1235  &  17.98(2)      &  300    &  37.1196  &  18.19(1)      &  300    &  37.1292  &  18.23(1)      &  600     \\
   37.3128  &  17.73(3)      &  300    &  37.3217  &  17.95(3)      &  300    &  37.3178  &  18.21(2)      &  300    &  37.3278  &  18.36(2)      &  600     \\
   38.2559  &  18.00(4)      &  300    &  38.2603  &  18.26(2)      &  300    &  38.2645  &  18.53(3)      &  300    &  38.2704  &  18.63(3)      &  600     \\
   39.2164  &  18.19(4)      &  300    &  39.2206  &  18.40(3)      &  300    &  39.2250  &  18.58(3)      &  300    &           &                &          \\
   40.0592  &  18.40(4)      &  480    &  40.0738  &  18.65(3)      &  480    &  40.0675  &  18.88(3)      &  480    &  40.0807  &  19.09(4)      &  600     \\
   45.2344  &  18.38(9)      &  300    &  45.2387  &  18.48(8)      &  300    &  45.2425  &  18.85(14)     &  300    &  45.2482  &  18.83(12)     &  600     \\
   46.0150  &  18.16(4)      &  500    &           &                &         &  46.0064  &  18.81(4)      &  600    &           &                &          \\
   47.1002  &  17.73(2)      &  720    &  47.2723  &  17.98(2)      &  720    &  47.1142  &  18.12(6)      &  720    &  47.2822  &  18.37(3)      &  900     \\
   48.0390  &  18.02(3)      &  600    &  48.1593  &  18.28(2)      &  900    &  48.0490  &  18.47(2)      &  900    &  48.1720  &  18.64(2)      & 1200     \\
   49.1236  &  18.43(5)      &  900    &           &                &         &  49.1764  &  18.79(4)      &  900    &           &                &          \\
   51.0626  &  17.85(2)      &  720    &           &                &         &  51.0714  &  18.20(1)      &  720    &           &                &          \\
   55.1715  &  18.70(16)     &  600      &           &                &         &  55.1920  &  19.17(6)    &  600    &           &                &          \\
   58.1532  &  18.53(4)      &  600    &           &                &         &  58.1606  &  19.01(2)      &  600    &           &                &          \\
 \hline
 \end{tabular}
\end{center}
\hspace*{1.0cm} {\em a} Times of mid integration (-2453500).
\end{minipage}
\end{table*}

\begin{figure}
\epsfig{file=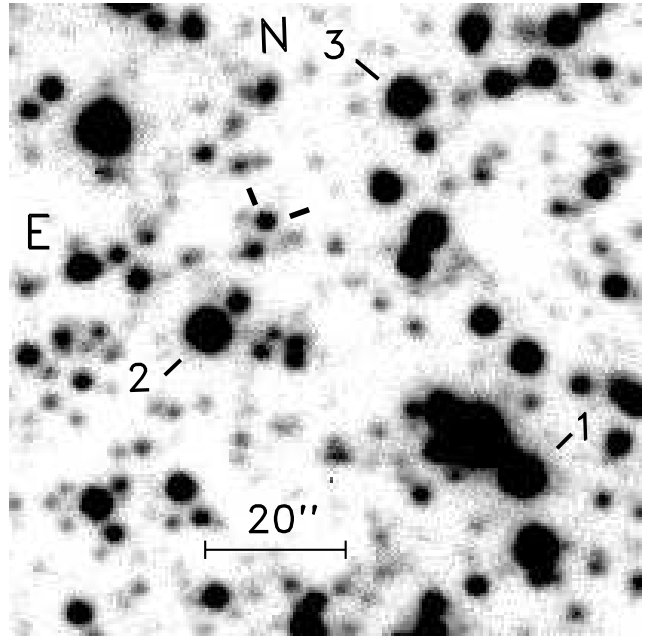, width=8.4cm } 
  \caption{A chart showing local standards used for SAX J1808.4-3658. This  
  image was taken when there was an {\it I\/} band excess on 2005 June 5 (HJD 2453527) 
  The source magnitude was  {\it I\/} = 16.72. The three local secondary standards listed in 
  Table 1 are marked with the numbers 1-3.  }
\end{figure}

\begin{figure}
\epsfig{file=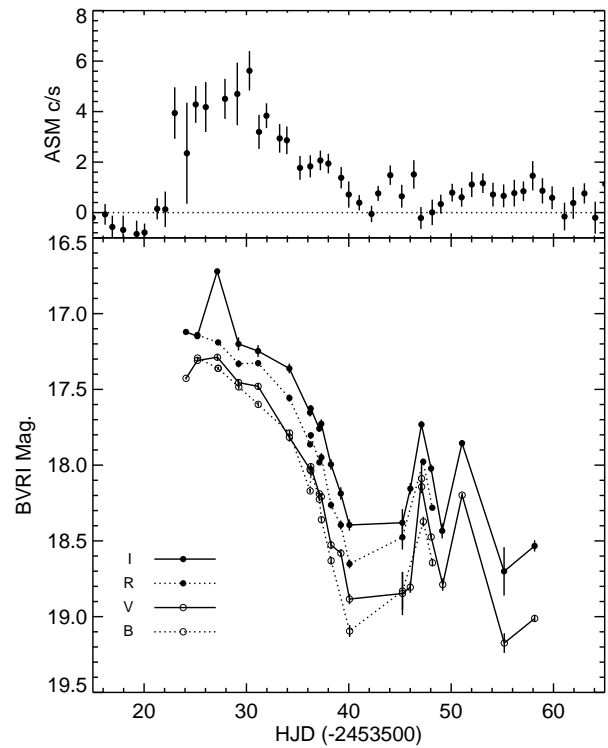, width=8.4cm }
  \caption{{\it RXTE} ASM light curve for SAX J1808.4-3658
  (upper panel) and {\it BVRI\/} band light curves (lower panel). } 
\end{figure}

\section{Results}
\subsection{The light curves}
The optical {\it BVRI\/} band and {\it Rossi X-ray Timing Explorer (RXTE)}, 
All Sky Monitor (ASM) (Levine et al. 1996) X-ray light curves 
are shown in Fig. 2. Both show a similar declining trend until about HJD 
3540 followed by two shorter duration flares similar to those 
seen in the 2000 and 2002 outbursts (Wachter et al. 2000; Wijnands et al. 2001, 
Wijnands, 2004).
There was also a large excess ($\sim 0.45$ mag) in the {\it I \/} 
band but not in the other colours on HJD 3527.

The decay time-scale during the first outburst (HJD 3522 to 3540) was 
$\sim 7$ days for both the optical and the X-ray emission. This is very 
similar to that during the discovery outburst in 1998 (Giles et al. 1999)
but much shorter than the $\sim 22 $ day time-scale for XTE J0929--314 (Giles
et al. 2005). Visual inspection of the light curves suggests that there is
a weak correlation between the optical 
and X-ray fluxes with the optical emission preceding the X-ray by a few
days as in the 1998 outburst (Giles et al. 1999). To test this hypothesis we converted the 
optical magnitudes to fluxes normalised to the peak ASM count rate and calculated the 
sum of the squares of the differences between the the two fluxes for delays varying between -10 and +10 days.
There is a broad minimum, for the whole dataset, corresponding to the optical emission leading the X-ray by -1 
to +3 days. When this analysis is restricted to the data during the secondary outbursts commencing at HJD 3542 
there is no significant minimum.  This lack of correlation may well be a consequence of inadequate 
sampling and the low statistical significance of the ASM data at these times.  

\subsection{Spectral changes}

In Fig. 3 we plot the {\it B-I\/} and {\it V-I\/} colour index changes during the 
observations. Where two measurements in the same colour were made on the same night 
we use the mean of the two. The solid line represents a linear fit to the data during the first 
steady decline in intensity ending at HJD 3540.  The anomalous point at HJD 3527 is discussed 
in section 3.3. Neither it nor the points measured after HJD 3540 are included in the linear fits.

Many of our nightly measurements were made consecutively and therefore
cluster within a binary phase duration of  $< 0.25$ but this was not always 
the case. In the absence of a continuous set of measurements we are unable to confirm 
the continued existence of the 0.12 mag binary modulation in {\it V\/} observed 
during the 1998 outburst by Giles et al. (1999).
Larger {\it VRI\/} modulations are reported during the quiescent state (Campana et al. 1998)
but no {\it B\/} modulation value is known for either state. We have taken the Campana et al. values 
(scaled down to the 1998 outburst {\it V\/} modulation), assumed a {\it B\/} modulation a 
little smaller than at {\it V\/}, and then used the precise ephemeris of Chakrabarty 
\& Morgan (1998) to examine the {\it B-I\/} and {\it V-I\/} shifts expected if any such 
modulations were actually present. We find that for {\it V-I\/} the introduced 
simulated modulation 'errors' are  $< 0.05$  mag. The scatter in the 
lower panel of fig 3 indicates that a modulation of this order may be present although we 
have no positive indication for it. For the {\it B-I\/} data the modulation 
'errors' are perhaps twice the size for {\it V-I\/} and of course rather dependent 
on the assumed {\it B\/} modulation. However, the small {\it B-I\/} scatter about the 
linear fit in the upper panel of fig 3 suggests that minimal modulation 
was actually present.

The most obvious features in both plots is the large increase due to the 
flare in {\it I\/} on HJD 3527 and two smaller but significant increases on HJD 3624 and 3546. We 
discuss these anomalies below. Overall {\it B-I\/} increased linearly
(became redder) until the intensity had decreased by about two magnitudes at the 
end of the primary outburst at HJD 3540.  This is illustrated by the solid lines in Fig. 3. The fact that the 
decline in emission is greatest in the blue end of the spectrum is suggestive of a viscosity driven 
``inside out'' transition leading to a decline in inner disc emission. 

During the secondary outburst peaking at HJD $\sim 3547$ the spectrum was at first significantly redder 
at HJD 3546 and thereafter {\it V-I\/} returned to the values ($0.4 \pm 0.1$) observed at the end of the 
initial intensity decline. Inspection of the light curves in Fig. 2 suggests that the increase in {\it V-I\/} on 
HJD 3546 was due to a brightening in {\it I\/} 
rather than a dimming in {\it V\/}. Another sudden change in colour occurred between the first and second 
nights of our observations (HJD 3524 and 3525).  These anomalous changes may be similar in nature to the 
{\it I\/} band flare on HJD 3527 but this cannot be confirmed in the absence of {\it R\/} \& {\it I\/} measurements 
at those times. An alternative explanation is that they are due to a mass 
transfer instability in which cool matter is dumped into the outer disc. This is followed by an increase in 
optical emission and colour temperature as the matter diffuses inwards.

\begin{figure}
\epsfig{file=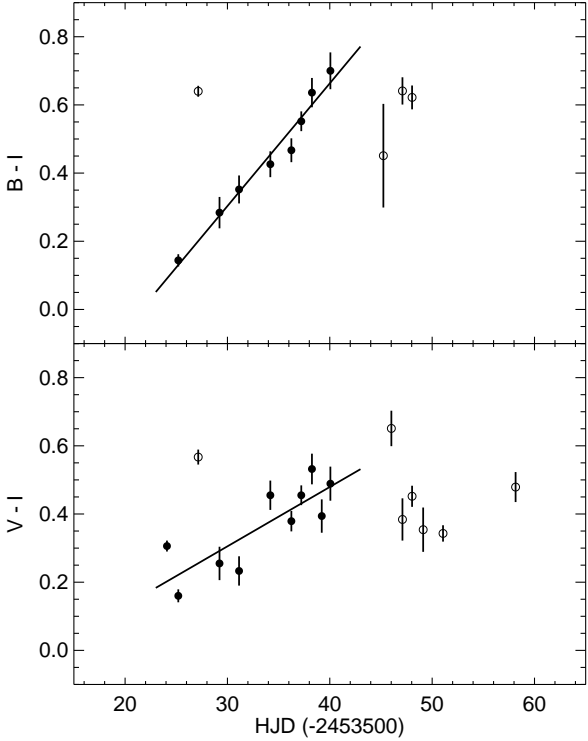, width=8.4 cm}
  \caption{Time dependence of the colour indices {\it B - I\/}  
  (upper panel) and  {\it V - I\/} (lower panel) for the SAX J1808.4--3658 observations. The solid 
  lines represent linear fits to the points marked by filled circles.} 
\end{figure}

On 14 occasions within Table 2 we have 4 colour {\it BVRI\/} 
measurements taken over a short interval (typically $\sim 1$ hr) on the same night.
Using these data and the bandwidth specifications for each filter we have derived
broadband {\it BVRI\/} spectra for these nights. In Fig. 4 we plot the spectra for
5 representative nights (HJD 3525, 3527, 3536, 3540 \& 3547) corresponding to the 
early bright state, the {\it I\/} band flare, the declining phase, the low intensity
phase and peak phase of the secondary outburst respectively. We also include the 
{\it V\/} \&{\it I\/} points taken on HJD 3546 during the early rising phase of the 
first secondary outburst.

\begin{figure}
\epsfig{file=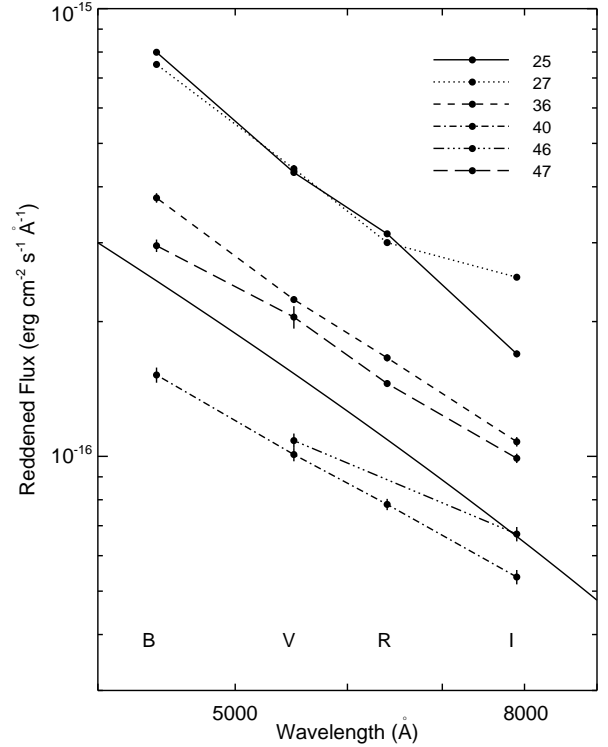, width=8.4cm }
  \caption{The {\it BVRI\/} broadband spectra for SAX J1808.4-3658. Each day's 
  spectrum is identified by the last two digits of the HJD of the observations. 
  The lines connect to the mean flux values 
  on each night. Some error limits are smaller than the points representing 
  the measurements. The dotted line represents a simple power law disc 
  emission model with exponent -3, arbitrary amplitude and interstellar 
  reddening corresponding to $A_{V}=0.68$. } 
\end{figure}

Also shown is a solid curve representing a power law approximation to the 
emission from an optically thick, X-ray heated disc. The distribution is 
given by the equation $F_{\lambda} \propto \lambda^{-3}e^{-A_{\lambda } 
/1.086}$ where $F_{\lambda }$ (Rieke \& Lebofsky  1985) is the reddened flux at wavelength $\lambda $ 
and $A_{\lambda }$ is the wavelength dependent reddening correction toward 
the source.  The spectrum is reddened assuming 
interstellar extinction $A_{V}=0.68$ (Wang et al. 2001). The amplitude is arbitrary.

The anomalous spectrum on HJD 3527 will be discussed in the next section. 
On most other nights the spectra are consistent with the reddened power law 
approximation expected for an X-ray heated accretion disc with interstellar 
extinction. The spectra became redder (cooler) with time as noted from the {\it B-I\/} colour index
plot (Fig. 3). A similar cooling trend was also apparent in broadband spectra 
taken during the 1998 outburst (Wang et al. 2001).  Overall the spectra 
were similar (spectral index $\sim 2.5-3.5$) to that in the 1998 outburst
(Wang et al. 2001)A somewhat steeper spectrum was reported for the 2003 outburst of the 
accretion-powered millisecond pulsar XTE J1814--338  (Krauss et al. 2005).

Except on the night of HJD 3527, there is no {\it I\/} band excess 
of the kind reported by Giles et al (2005) for XTE J0929--314 although, in the absence of {\it B\/} 
\& {\it R\/} measurements, we cannot exclude the possibility that there was a single colour {\it I\/} 
band excess on HJD 3524 and in the early phase of the secondary outburst at HJD 3546.

\subsection{{\it I\/} band flare}

On the night of 2005, June 5 the {\it I\/} band measurement at HJD 3527.1449 
revealed a $\sim 0.45$ magnitude excess above the trend from previous and succeeding 
nights. There was no evidence of any excess above the trend-lines for any of the other
colours all of which were measured within $\sim 2$ hrs of the {\it I\/} band. The 
existence of an {\it I\/} band excess is also obvious in the broadband 
spectrum for the night as shown in Fig. 4.  

The absence of any perceptible increase in {\it R,\ V\/} or {\it B\/} indicates that the enhanced 
radiation in {\it I\/} was sharply cut-off in wavelength or was of duration less
than 32 minutes when the next measurement (in {\it V\/}) was made. An enhanced orbital modulation can be 
ruled out since there was no increase in {\it R\/} or {\it B\/} measured one orbital cycle 
after {\it I\/}.

{\it RXTE} Proportional Counter Array (PCA) measurements show that a typical Type I X-ray burst occurred 
almost in the middle of  our 300 sec {\it I\/} band integration (private 
communication, Wijnands \& Klein-Wolt). The peak amplitude was $\sim 40$ times the baseline flux and 
the duration (to $ 5\% $ of baseline) was $\sim 35$ s. The associated optical burst generated by 
reprocessing of the burst X-rays undoubtedly contributed to the increase in {\it I\/}. 

Type I X-ray bursts have been observed from many neutron star binaries and typically last 
10--20 sec although a few have lasted 
up to 150 seconds ( Kong et al. 2000). For the few examples available with simultaneous X-ray 
and optical data, mostly from 4U 1636--53 (Pedersen et al. 1982; Lawrence et al. 1983; 
Matsuoka et al. 1984) and GS 1826--24 ( Kong et al. 2000), the optical and X-ray profiles are similar 
in shape and duration  There is usually evidence for an optical lag of a few seconds corresponding to 
the light travel time to the reprocessing site. 

Preliminary calculations suggest that reprocessed radiation contributed only a small fraction of the 
increase seen in the {\it I\/} band. A paper on this event, combining the X-ray, optical and radio data, 
is in preparation. This will include detailed calculations setting limits on the 
optical flux expected from reprocessed radiation. For now we rely on approximate
estimates based on scaling arguments using published simultaneous optical and X-ray observations 
of the X-ray burst source 4U 1636--53. The peak amplitudes of the bursts were $\sim 40 $
times the baseline X-ray flux in both systems although the duration of the SAX J1808.4--3658
burst was about twice as long as typical bursts seen in 4U 1636--53.  

Optical bursts are generated by reprocessing of burst X-rays incident on the disc and the companion.
The fraction observed from the disc is expected to be strongly dependent on the inclination of the system. 
The component coming from the companion star will also be dependent on on the inclination (though not 
so strongly) and will be proportional to the solid angle subtended by the companion Roche lobe at the burst 
source. Its amplitude will depend on the orbital phase at which the burst occurred. We assume in the calculations 
below that the fraction arising from the companion is equal to the fractional orbital modulation observed during 
transient outbursts.

The inclination angles are poorly known in both SAX J1808.4--3658 and 4U 1636--53 but  are believed 
to be $\sim 60 \degr $ in both systems (Chakrabarty \& Morgan 1998; Homer et al. 2002; Frank, 
King \& Lasota 1987). The 
X-ray burst occurred at binary phase $\phi = 0.08$  (phase zero corresponds to the time when 
the companion star is at its maximum distance from the observer), almost optimum time for observing reprocessed 
radiation from the companion. Using the estimated dimensions of the two systems (Frank, 
King \& Lasota 1987; Chakrabarty \& Morgan 1998) we find that the solid angle of the Roche lobe
in the  4U 1636--53 system is $\sim 4$ times that in  SAX J1808.4--3658. Hence the peak optical burst 
flux from the companion in 4U 1636--53 will be 4 times that in SAX J1808.4--3658 assuming that the X-ray 
bursts are similar in peak flux. Orbital modulation measurements show, however, that the fraction of 
radiation reprocessed on the companion to 4U 1636--53 is $ \sim 25$ per cent (Giles et al. 2002) which 
is only twice that in SAX J1808.4--3658 (Giles et al. 1999). Hence, for the same 
peak X-ray burst flux, the observed reprocessed radiation from the disc is also larger in 4U 1636--53 
than in SAX J1808.4--3658.  This may be a consequence of a larger disc size or perhaps of a smaller 
inclination angle for 4U 1636--53. The two conditions above are satisfied if we assume that, for the 
same peak X-ray burst flux, optical bursts in 4U 1636--53 are twice as large as those in SAX J1808.4--3658.

The typical peak optical burst flux to baseline ratio in 4U 1636--53 is $\sim 1.5$ 
(Pedersen et al. 1982; Lawrence et al. 1983). Bursts in GS 1826--24 are much longer in 
duration but the ratio of optical to X-ray burst height is less (Kong et al. 2000). 
Given that the X-ray bursts in SAX J1808.4--3658  and 4U 1636--53 are of similar peak amplitude 
we expect the optical burst in SAX J1808.4--3658 to have half the  relative amplitude ($\sim 0.75$ times 
baseline). The total {\it I\/} band integration 
time (300 s) was much longer than the burst duration and this will reduce the amplitude of 
the reprocessing signal by the ratio of the integrated optical burst flux to the 'normal' optical flux. 
This ratio is $\sim 0.5\times20/300$ assuming the optical burst profile was triangular with effective duration 
$ \sim 20$ s. Hence the increase in the integrated {\it I\/} band flux due to reprocessing 
would be $\sim 2.5$ per cent above baseline. This is a factor 20 less than 
the observed $\sim 50 $ per cent increase in {\it I\/}. There are many uncertainties in the above argument but 
it seems unlikely that all the {\it I\/} band excess can be due to reprocessed X-ray burst emission.  
The burst may have been the trigger for an on-going synchrotron emission event.

Rupen et al. (2005) detected weak 4.86 and 8.46 GHz radio emission from SAX J1808.4--3658
on 2005 June 7, 11 \&16 (HJD 3529, 3533 \& 3538) and suggested it was due to synchrotron emission.  
Rea et al. (2005) measured {\it V,\ R\/}  \& {\it I\/} magnitudes on 2005 June 5 (HJD $ \sim 3526.5 $)
and set a $5\sigma$ upper limit of 16.5 on {\it H\/} band IR emission. Their measurements of the 
optical magnitudes are of low precision but are more consistent
with our 'normal' spectrum of June 3 than with the 'anomalous' data of June 5. Their {\it H\/} band upper
limit is slightly above an extrapolation of the 'normal' spectrum but is more consistent with it than with 
the 'anomalous' one. The measurements by Rea et al. (2005) were made $\sim 12$ hrs before our 
observations of June 5. It seems likely that the IR excess was not present at that time and we note also that no 
radio emission was detected on 2005 June 4 the day before our observation
(Rupen et al. 2005). We conclude that the {\it I\/} band enhancement was due, at least in part, to synchrotron 
emission. It commenced less than 12 hours before our observation and was probably triggered by the 
type I X-ray burst. Enhanced {\it I\/} band emission lasted less than 60 hours 
since it was not present on June 7 (HJD 3529). This implies also that the synchrotron spectrum cut-off frequency decreased since radio synchrotron emission was detected on that day and continued for at least another 9 days 
(Rupen et al. 2005). 

Giles et al. (2005) reported a variable {\it I\/} band excess in another
accreting millisecond pulsar, XTE J0929--314, and suggested that the anomalous 
optical and IR spectra in SAX J1808.4--3658 reported by Wang et al. (2001) may be similar in 
nature. Krauss et al. (2005) observed 
an {\it I\/} band excess in a third accreting millisecond pulsar XTE J1814--318 and suggested
that it was probably due to synchrotron emission. 
Hence we conclude that variable synchrotron emission is common in these systems.

\section{Discussion}
Our observations show that on most nights the optical spectra in outburst were consistent with
emission from an X-ray heated accretion disc.  The
spectra became redder as the intensity decreased. A similar decrease in colour
temperature was also apparent in the 1998 outburst of this source (Wang et al. 2001). 
These changes  are consistent with an 
'inside-out' transition in the disc i.e. one triggered by a viscosity change driven instability
in the inner disc. In this scenario the $\sim 7$ day time-scale for the outburst decline corresponds to
the time-scale for depletion of matter in the inner disc.

\begin{figure}
\epsfig{file=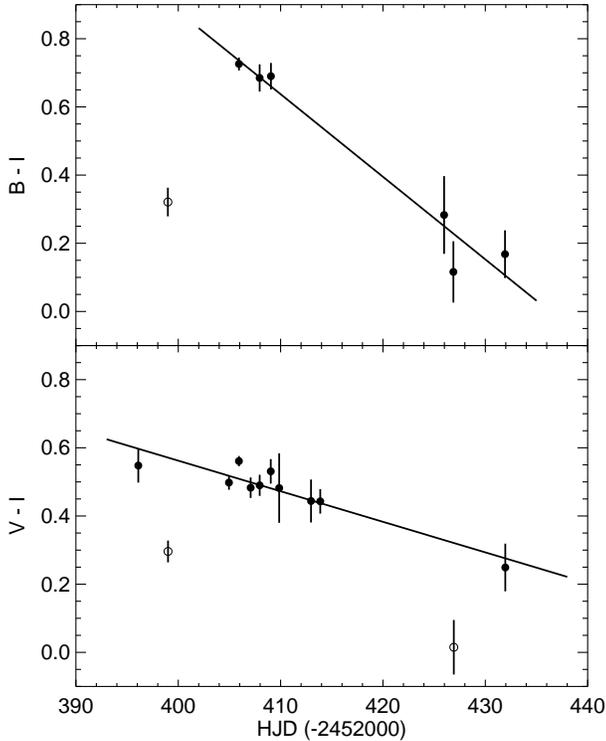, width=8.4 cm}
  \caption{Time dependence of the colour indices {\it B - I\/} (upper panel) and {\it V - I\/} 
  (lower panel) for the accreting millisecond pulsar XTE J0929--314 measured by Giles et al. (2005). The 
  solid lines represent linear fits to the points marked by filled circles.}
\end{figure}

The spectral changes in this system are remarkably different from that in the accreting millisecond 
pulsar XTE J0929--314. In Fig.5 we show the time dependence of {\it B-I\/} and {\it V-I\/} during 
the 2002 outburst in XTE J0929--314 (Giles et al. 2005). These are drawn to the same scale as in Fig. 3.
Again we note the strong influence of the changes in {\it B\/} (predominately inner disc emission) 
on the spectrum. The change in spectral colour is, however, opposite in direction to that seen in SAX 
J1808.4--3658 suggesting different outburst mechanisms in the two systems. In XTE J0929--314 the disc became 
hotter as the intensity decreased during the 2002 outburst (Giles et al. 2005). This is suggestive 
of a mass transfer (outside-in) instability. It is not clear why these systems should differ in this way
although we note that the companion in XTE J0929--314 is thought to be a degenerate helium core with
remnant envelope and a mass $\sim 0.01M_\odot $ (Galloway et al. 2002). The companion to SAX J1808.4--3658 
is a brown dwarf at least 5 times as massive (Bildsten \& Chakrabarty, 2001). This may affect the 
stability of the atmospheres of the companions while undergoing X-ray heating. We note however that a
mass transfer instability may have contributed to the quasi-periodic secondary outbursts in 
SAX J1808.4--3658 commencing at HJD $\sim 3545$ during our observation. 

On several occasions we have detected highly significant changes from night to night in the colour 
indices suggestive of sudden changes in disc structure. These were particularly strong in {\it B - I\/}. 
Giles et al. (2005) reported similar anomalous changes in XTE J0929--314. During future outbursts four 
colour ({\it BVRI\/}) sampling several times per night will be required to adequately characterise 
the evolution of these changes.

The {\it I\/} band excess seen on 2005 June 5 was due in part to an optical burst generated by 
reprocessing of the X-rays from a Type I burst which occurred during our {\it I\/} band integration. 
Preliminary calculations indicate, 
however, that reprocessing alone is insufficient to explain the observed increase. We suggest that the X-ray burst 
triggered synchrotron emission which had disappeared at optical wavelengths within two days but 
which persisted at radio frequencies until at least 2005 June 16 (Rupen et al. 2005). If this scenario is correct 
the increase in {\it I\/} band flux due to synchrotron radiation was at least twice the normal flux since synchrotron 
emission did not commence until after the middle of the exposure.  We are not aware of 
any previous evidence for synchrotron emission being triggered by an X-ray burst. 

Of the seven known accreting millisecond pulsars, three (SAX J1808.4--3658, 
XTE J0929--314 \& XTE J1814--338) are now known to have transient near IR emission.
It seems likely that this is due to synchrotron emission which extends at times to {\it I\/} band
wavelengths. Synchrotron emission at radio and IR wavelengths has been detected from 
a number of other X-ray binaries during outburst (Fender, 2001) but it seems it is particularly
common in accreting millisecond pulsars.

\section{Acknowledgments}
This research has made use of data obtained through the High Energy Astrophysics Science 
Archive Research Center Online Service, provided by the NASA / Goddard 
Space Flight Center. We are grateful to Rudy Wijnands, Marc Klein-Wolt and Manu Linares for 
providing us with unpublished {\it RXTE} PCA data. 
We thank Rossi Corrales for assistance with the observations, 
Kym Hill  and Stefan Dieters for helpful comments and  gratefully acknowledge 
financial support for the Mt Canopus Observatory by David Warren. ABG thanks the University of 
Tasmania Antarctic CRC for the use of computer facilities.

\label{lastpage}

\end{document}